\documentclass[journal]{IEEEtran}  

\usepackage{cite}
\usepackage{graphics}
\usepackage{epsfig}
\usepackage{mathptmx} 
\usepackage{times} 
\usepackage{amsmath}
\usepackage{amssymb} 
\usepackage{amsthm} 
\usepackage{mathptmx}
\usepackage{dblfloatfix} 
\usepackage{comment}
\usepackage{mathtools}
\usepackage{tabularx} 
\usepackage{multicol}
\usepackage{xcolor}

\usepackage[section]{placeins}
\usepackage{cuted}

\usepackage{tabularray}
\usepackage{float}
\usepackage[shortlabels]{enumitem}
\usepackage{placeins}
\DeclareMathAlphabet{\mathcal}{OMS}{cmsy}{m}{n}

\theoremstyle{remark}

\theoremstyle{definition}

\usepackage[utf8]{inputenc}
\usepackage[english]{babel}

\usepackage{multirow}
\usepackage{tabularx}
\usepackage{adjustbox} 

\newcommand{\PreserveBackslash}[1]{\let\temp=\\#1\let\\=\temp}
\newcolumntype{C}[1]{>{\PreserveBackslash\centering}p{#1}}
\newcolumntype{R}[1]{>{\PreserveBackslash\raggedleft}p{#1}}
\newcolumntype{L}[1]{>{\PreserveBackslash\raggedright}p{#1}}

\newcommand{\nrc}[1]{\textcolor{black}{#1}}

\begin{document}

\title{An Approach to Evaluate Modeling Adequacy for Small-Signal Stability Analysis of IBR-related SSOs in Multimachine Systems
}

\author{Lilan~Karunaratne,~\IEEEmembership{Student Member,~IEEE,}
        Nilanjan~Ray~Chaudhuri,~\IEEEmembership{Senior Member,~IEEE,}
        Amirthagunaraj~Yogarathnam,~\IEEEmembership{Member,~IEEE},
        and~Meng~Yue,~\IEEEmembership{Member,~IEEE \vspace{-20pt}}
\thanks{This work is supported under Agreement 37532 by the Advanced Grid Modeling (AGM) Program of the Office of Electricity, the Department of Energy (DOE).}
\thanks{L. Karunaratne and N. R. Chaudhuri are with the School of Electrical Engineering and Computer
Science, Penn State University, University Park, PA, USA e-mail: lvk5363@psu.edu, nuc88@psu.edu.}
\thanks{A. Yogarathnam and M. Yue are with the Interdisciplinary Science Department, Brookhaven National Laboratory, Upton, NY, USA e-mail: ayogarath@bnl.gov, yuemeng@bnl.gov.}
}

\maketitle

\begin{abstract}
Time-varying phasor-based analysis of subsynchronous oscillations (SSOs) involving grid-following converters (GFLCs) and its benchmarking with electromagnetic transient (EMT) models  have so far been restricted to highly simplified grid models with constant voltage sources behind series $R$-$L$ circuits. In this paper, modeling adequacy of bulk power systems with synchronous generators (SGs), transmission systems, loads, and GFLCs are considered. To this end, we revisit the notions of time-varying phasor calculus, highlighting the distinction between space-phasor-calculus (SPC) and two often interchangeably used frameworks namely baseband-$abc$ and generalized averaging. We present the models of grids in SPC framework that include transmission line dynamics, load dynamics, and SG stator transients. Next, we propose a generic approach to study modeling adequacy in small-signal sense by (a) identifying critical modes through eigenvalue and singular value analysis followed by (b) using weighted maximum singular value error magnitudes as metrics, and (c) further cross-validation. Using a modified $4$-machine IEEE benchmark model with up to $3$ GFLCs we show that SPC framework can be used for analysis of SSOs. Further, we consider the quasistationary phasor calculus (QPC) framework that neglects transmission line, load, and SG stator dynamics to show its adequacy in SSO modeling and analysis. 
Time-domain and frequency-domain results with EMT models are also presented. 
\end{abstract}



\IEEEpeerreviewmaketitle

\vspace{-10pt}
\section{Introduction}\label{sec:Intro}
\IEEEPARstart{P}{roliferation} of inverter-based resources (IBRs) in today's power system has pushed it into an unfamiliar territory. IBRs function based on two basic types of converters - (a) grid-following converters (GFLCs) and (b) grid-forming converters (GFCs). Among these, the first one is commonly adopted whereas GFC represents a relatively new technology under grid-connected operation. In the recent past, bulk power systems in different parts of the world have witnessed subsynchronous oscillations (SSOs) involving GFLCs. The IEEE Power \& Energy Society (PES) IBR SSO task force has compiled a list of $19$ such events \cite{IBRSSOTF}. Root cause analysis fundamentally divides such phenomena into three classes $-$ (a) series capacitor SSO, (b) weak grid SSO, and (c) inter-IBR SSO. A comprehensive list of literature in this area can be found in \cite{IBRSSOTF}, whereas the possibility of inter-IBR SSO was shown in \cite{Lingling-InterIBR}. \nrc{Understanding adequacy of dynamic models used for planning studies to investigate SSOs stemming from GFLC-based IBRs needs urgent attention as it will take some time before GFC-based IBRs are integrated to the system. Adequacy analysis in presence of GFC-based IBRs will be considered separately and is outside the scope of this paper. }
\vspace{-5pt}
\subsection{Motivation behind studying modeling adequacy}
With this background, it is obvious that a clear understanding of the interaction of IBRs with the rest of the components in the grid is required as the IBRs start dominating the generation portfolio. This underscores the importance of studying modeling adequacy of the traditional quasistationary assumption of transmission networks in planning models. \textit{More specific, we focus on the modeling adequacy of balanced systems with SSOs stemming from IBRs that are based on GFLC technology.} 

We note that modeling complexity increases significantly if transmission network dynamics and synchronous generator (SG) stator transients need to be considered. This also demands much shorter integration time steps for time-domain simulations and will be prohibitively difficult to simulate without variable time-step integration methods. Modeling adequacy study focused on SSOs for a particular system can reveal if neglecting such dynamics is acceptable. It is therefore crucial to perform such evaluations prior to conducting exhaustive time-domain simulations for planning studies. 
\vspace{-5pt}
\subsection{Literature review} 
Literature on modeling adequacy of IBR-dominated systems in the context of SSOs is quite limited. For example, most of the work in this area \cite{Lingling-InterIBR,Lingling-19-Type4WindModel,Lingling-21-ReducedAnalyticalPV,Lingling-23-NewIBRoscType,SSO_ibr} have considered a highly simplified model of the power grid that is represented by an ideal voltage source behind a series $R$-$L$ circuit. The dynamics of the $R$-$L$ circuit has been taken into account without any solid justification.  

Although not specific to SSOs, in \cite{Strunz-21-QPC-DPC} a five-step methodology for determining modeling adequacy of grids with IBRs was presented. These steps are -- (1) dynamic modeling, (2) frequency response analysis, (3) modal analysis, (4) sensitivity analysis, and (5) validation through time-domain simulation. The paper shows that quasistationary phasor calculus (QPC)-based models can produce inaccurate results compared to dynamic phasor calculus (DPC)-based models in presence of GFLCs, GFCs, and SGs. 

Notwithstanding the importance of \cite{Strunz-21-QPC-DPC}, the paper has the following gaps --
(1) \textit{Modeling:} The paper does not clearly articulate the DPC-based modeling framework when it comes to interfacing IBR models and SG models including stator transients, with the lumped parameter-based dynamic model of the transmission network; given different frames of references needed in the process. Moreover, SG models including stator transients are not commonly used and therefore need to be clearly presented.
(2) \textit{Frequency response analysis:} The authors proposed to compare Bode magnitude plots to determine frequency threshold $f_*$ up to which models match within a $\%$ tolerance. However, such \textit{point-by-point} error calculation may give misleading information since there could be a large $\%$ deviation at a particular frequency where the gain is not high. Moreover, such modeling mismatch should consider the multi-input-multi-output (MIMO) nature of the system for which singular value plots are more appropriate. The authors in \cite{Strunz-21-QPC-DPC} did consider the $\mathcal{H}_\infty$ norm to conclude if a model is going to produce substantially different response from the other. However, the $\mathcal{H}_\infty$ norm gives the maximum singular value of a MIMO system over the entire frequency range. For two models, the values of this norm can be very close to each other, but they may correspond to two different frequencies. Therefore, it should not be considered as a metric on a stand alone basis.  

For SSO studies, time-varying phasor models have been considered in the literature \cite{Lingling-InterIBR,Lingling-19-Type4WindModel,Lingling-21-ReducedAnalyticalPV,Lingling-23-NewIBRoscType}. In this context, the aspect that requires further clarity is the interchangeable usage of \textit{baseband $abc$-frame representation} and the \textit{time-varying dominant Fourier coefficient-based representation} as `dynamic phasor' (DP). The first paper that developed a comprehensive calculus of the baseband $abc$-frame representation for bulk power system model was \cite{Mani-94-DynamicPhasor}, which is also followed in \cite{Strunz-21-QPC-DPC}. On the other hand, the theory of generalized averaging proposed in \cite{Verghese-91-DynamicPhasor} led to the second representation, which has widely been called DPs in papers including \cite{Stankovic-00-DynamicPhasor}, \cite{Bozhko-16-DynamicPhasor}, and many others. As pointed out in \cite{Mani-94-DynamicPhasor}, it is essential to limit the bandwidth of the phasors to strictly below the carrier frequency ($50$ or $60$ Hz in power systems), which however is not a constraint for a Park's transformation-based approach proposed in \cite{Mani-95-DQ0} that utilizes space-phasor-based calculus (SPC) in rotating $d$-$q$ frame. The above discussion shows that some clarity is needed regarding the common features and differences among the multitude of approaches describing the time-varying phasors.
\vspace{-5pt}
\subsection{Contributions of this work}
This paper presents\\ 
(1) a comparative summary that clarifies different notions of the so-called \textit{time-varying phasors} that exist in literature. It highlights that for a balanced system, SPC in $dq0$ frame does not involve any approximation, whereas baseband-$abc$ is restricted in phasor speed and generalized averaging involves approximation due to truncation of Fourier coefficients.\\
(2) SPC- and QPC-based modeling approaches in rotating $d$-$q$ frames for a generic power system with SGs, GFLCs, transmission network and loads with sufficient clarity.\\
(3) a generic approach to study adequacy of MIMO models in small-signal sense that involves identifying critical modes through eigenvalue and singular value analyses followed by using weighted maximum singular value error magnitudes as adequacy metrics.\\
(4) case studies on a modified IEEE $4$-machine test system to demonstrate the applicability of the proposed method for evaluating modeling adequacy followed by stability analysis. Time-domain and frequency-domain results with EMT models are also presented.

\section{Different Representations of Time-Varying Phasor Calculus}\label{sec:TimeVarPhasor}
The notion of time-varying phasor calculus has taken three different forms, which are elaborated below. Please see Chapter $3$ of \cite{Demiray2008} for a review on this topic. 

Although these concepts are not new, there are misconceptions that led to interchangeable usage of these three forms, which have not been comprehensively clarified in literature. Our goal is to fill this gap.

\textit{1. Baseband-$abc$ representation:} This representation is first proposed in \cite{Mani-94-DynamicPhasor}, where a \textit{modulated single-phase} signal $x(t) = X(t)cos(\omega_s t + \theta(t)) \in \mathcal{B}$ can be mapped into a time-varying phasor of the form $x_{bb}(t) = X(t)e^{j\theta(t)} \in \mathcal{L}$, where $\omega_s$ is the synchronous speed in electrical rad/s. The phasor operator $\Upsilon$ can be defined as a mapping $\Upsilon: \mathcal{B} \rightarrow \mathcal{L}$ such that $x_{bb}(t) = \Upsilon(x(t))$ and $x(t) = \Re{\{x_{bb}(t)e^{j\omega_s t}\}},~\forall t \in \mathbb{R}$, where, $\Re\{\cdot\}$ is the real part operator. The operator $\Upsilon$ is essentially a composition of transforming the modulated signal $x(t)$ to the analytic signal $x_a(t) = x(t) + j\mathcal{H}\{x(t)\}$ and the frequency shift operation leading to $x_{bb}(t) = x_a (t)e^{-j\omega_st}$, where $\mathcal{H}\{\cdot\}$ denotes the Hilbert transform. \vspace{-5pt}

\begin{figure}[h!]
    \centerline{    \includegraphics[width=0.4\textwidth]{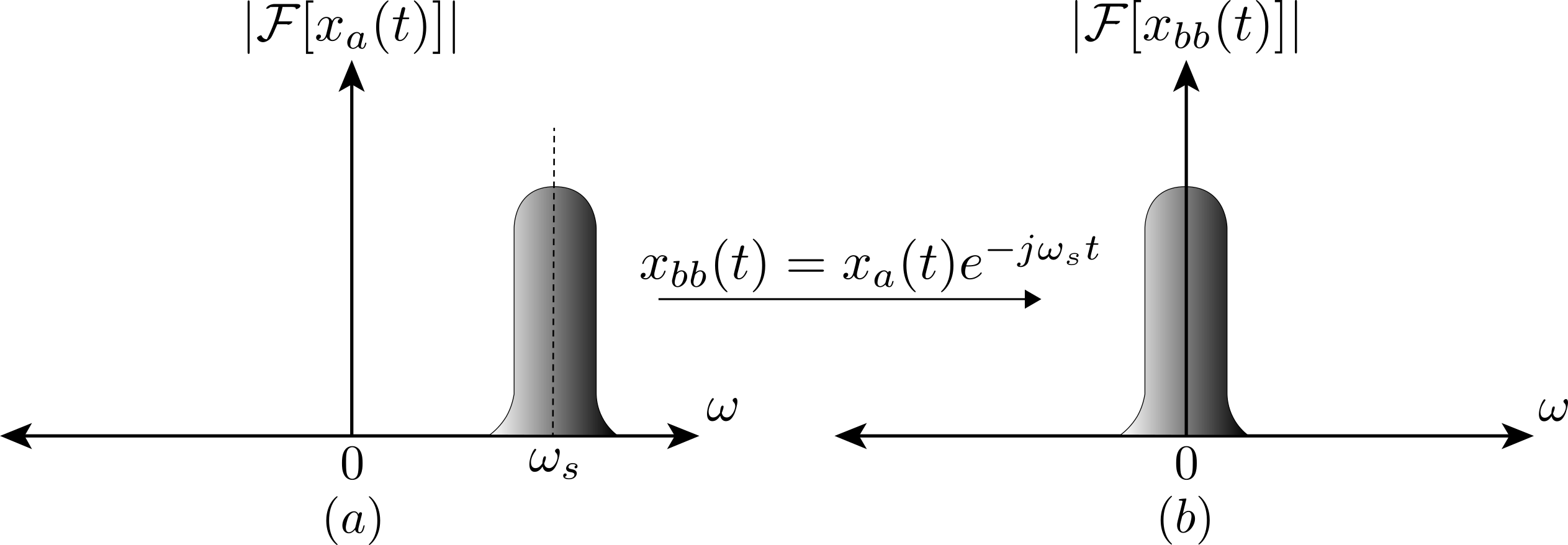}}
    \vspace{-8pt}
    \caption{Signal spectrum of (a) an analytic and (b) a baseband signal.}
    \label{fig:spectrum}
\end{figure} 
\vspace{-5pt}

The following properties of $\Upsilon$ are essential for its practical application in power systems.\\
\textit{$\Upsilon: \mathcal{B} \rightarrow \mathcal{L}$ should be (a) bijective, and (b) linear.}\\
The first property ensures that the phasor operator is well-defined and the second property is essential for satisfying KCL, KVL, and power balance of the network in time-varying phasor domain. 

It was established in \cite{Mani-94-DynamicPhasor} that these properties will hold if $\mathcal{L}$ is restricted to the set of low-pass phasors and as a consequence $\mathcal{B}$ becomes the set of band-pass signals as dictated by $x(t) = \Re{\{x_{bb}(t)e^{j\omega_s t}\}},~\forall t \in \mathbb{R}$. In other words, $\mathcal{F}\{x_{bb}(t)\} = \mathbf{X}_{bb}(j\omega) = 0, \omega \geq \omega_s ~\text{and} ~\omega \leq -\omega_s$ and $\mathcal{F}\{x(t)\} = \mathbf{X}(j\omega) = 0, \omega = 0 ~\text{and} ~\omega \geq 2\omega_s$, where $\mathcal{F}\{\cdot\}$ is the Fourier transformation. These constraints are shown in Fig.~\ref{fig:spectrum}. This representation can be applied to individual phases of the power system for balanced condition as well as unbalanced condition using sequence transformation.

Finally, for $x(t), \frac{\mathrm{d} x(t)}{\mathrm{d} t} \in \mathcal{B}$ the derivative operation satisfies the following relationship \vspace{-2pt}
\begin{equation}
\Upsilon\left ( \frac{\mathrm{d} x(t)}{\mathrm{d} t} \right ) = \frac{\mathrm{d} x_{bb}(t) }{\mathrm{d} t} + j\omega_s x_{bb}(t).
\end{equation}

\textit{2. Space-phasor-based representation in $dq0$ frame:} The space-phasor (also called space-vector) $\bar{x}(t)$ is defined as the transformation of arbitrary three-phase signals $x_a(t)$, $x_b(t)$, and $x_c(t)$ such that $\bar{x}(t) = x_\alpha (t) + jx_\beta (t) = \frac{2}{3}\begin{bmatrix}
1 & \alpha & \alpha^*
\end{bmatrix}\textbf{x}(t)$, where $\alpha = e^{j\frac{2\pi}{3}}$, $(.)^*$ denotes conjugate operation, and $\textbf{x}(t) = [x_a(t)~ x_b(t) ~x_c(t)]^T$. In other words, this transformation leads to \textit{orthogonalization} of the $abc$ frame quantities. If $x_a(t) + x_b(t) + x_c(t) \equiv 0$, then $\bar{x}(t)$ holds the instantaneous information of $\textbf{x}(t)$.

Of particular interest is the special case of \textit{balanced} set of \textit{modulated} signals \cite{Mani-95-DQ0} of the form $x_a(t) = X(t)cos(\omega_s t + \theta(t))$, $x_b(t) = X(t)cos(\omega_s t + \theta(t) - \frac{2\pi}{3})$, and $x_c(t) = X(t)cos(\omega_s t + \theta(t) - \frac{4\pi}{3})$ for which $\bar{x}(t) = X(t)e^{j(\omega_s t + \theta(t))}$. The time-varying phasor $\bar{x}_{bb}(t)$ is calculated by the frequency shift operation on $\bar{x}(t)$ leading to $\bar{x}_{bb}(t) = x_d(t) + jx_q(t) = \bar{x}(t)e^{-j\rho(t)}$, where $\frac{\mathrm{d} \rho (t) }{\mathrm{d} t} = \omega (t)$. The operator $\bar{\Upsilon} : \mathcal{M} \rightarrow \mathcal{D}$ is defined as $\bar{x}_{bb}(t) = \bar{\Upsilon}(\textbf{x}(t))$, where $\mathcal{M}$ is a vector space representing the set of balanced three-phase signals. It can be shown that 
$\bar{\Upsilon}(\textbf{x}(t)) = [1~j~0] \mathcal{P}(t)\textbf{x}(t)$, where $\mathcal{P}(t) = \frac{2}{3}\begin{bmatrix}
\cos\rho(t) & \cos\left ( \rho(t) - \frac{2\pi}{3} \right ) & \cos\left ( \rho(t) + \frac{2\pi}{3} \right ) \\ 
-\sin\rho(t) & -\sin\left ( \rho(t) - \frac{2\pi}{3} \right ) & -\sin\left ( \rho(t) + \frac{2\pi}{3} \right ) \\ 
\frac{1}{2} & \frac{1}{2} & \frac{1}{2}
\end{bmatrix}$ 
is the Park's transformation matrix. 

Notice that $\bar{\Upsilon}(\textbf{x}(t))$ is \textit{a bijective linear transformation} since $\mathcal{P}(t)$ is invertible. Finally, for $\textbf{x}(t) \in \mathcal{M}$ the derivative operation satisfies the following relationship \vspace{-3pt}
\begin{equation}  \vspace{-2pt}
\bar{\Upsilon}\left ( \frac{\mathrm{d} \textbf{x}(t)}{\mathrm{d} t} \right ) = \frac{\mathrm{d} \bar{x}_{bb}(t) }{\mathrm{d} t} + j\omega(t) \bar{x}_{bb}(t).
\end{equation} 
Also, if $\omega(t) = \omega_s$ is chosen as in \cite{Mani-95-DQ0}, then $\bar{x}_{bb}(t) = x_d(t) + jx_q(t) = \bar{x}(t)e^{-j\omega_s t} = X(t)e^{j\theta(t)}$, which is similar to the baseband-$abc$ case. 

\textit{3. Generalized averaging theory-based representation:} In \cite{Verghese-91-DynamicPhasor} the generalized averaging theory was proposed, which expresses a near-periodic (possibly complex) time-domain waveform $x(\tau)$ in the interval $\tau \in (t - T, t]$ using a Fourier series of the form $x(\tau) = \sum_{k = -\infty}^{\infty} X_k(t) e^{jk\omega_s\tau}$, where $\omega_s = \frac{2\pi}{T}$, $k \in \mathbb{Z}$, and $X_k(t)$ are the complex Fourier coefficients that vary with time as the window of width $T$ slides over the signal. The $k$th coefficient, also called the \textit{$k$th phasor}, can be determined at time $t$ by the following \textit{averaging} operation $X_k(t) = \frac{1}{T}\int_{t-T}^{t} x(\tau) e^{-jk\omega_s\tau}d\tau = \left \langle x \right \rangle_k (t)$.

The derivative operation satisfies the following relation
\begin{equation}
    \left \langle \frac{\mathrm{d} x}{\mathrm{d} t} \right \rangle_k (t) = \frac{\mathrm{d} \left \langle x \right \rangle_{k}(t)}{\mathrm{d} t} + jk\omega_s\left \langle x \right \rangle_{k}(t)
\end{equation}
Note that the time-domain waveform $x(\tau)$ above can be $abc$ phase quantities or $dq0$ frame quantities. 

In \cite{Stankovic-02-DP}, the generalized averaging method was applied to the three-phase case to determine the $k$th dynamical +ve, -ve, and 0-sequence components $\left \langle x \right \rangle_{p,k}(t)$, $\left \langle x \right \rangle_{n,k}(t)$, and $\left \langle x \right \rangle_{z,k}(t)$, respectively, in the following form $\begin{bmatrix}
\left \langle x \right \rangle_{p,k}(t) & \left \langle x \right \rangle_{n,k}(t)  & \left \langle x \right \rangle_{z,k}(t)
\end{bmatrix}^T = \frac{1}{T} \int_{t-T}^{t}e^{jk\omega_s\tau}\mathcal{T}^H 
\textbf{x}(\tau)d\tau$, where, $\mathcal{T} = \frac{1}{\sqrt{3}}\begin{bmatrix}
1 & 1 & 1\\ 
\alpha^* & \alpha & 1\\ 
\alpha & \alpha^* & 1
\end{bmatrix}$ and $(.)^H$ denotes Hermitian operation.

In this approach we are interested in a good approximation provided by the set $\mathcal{U}$ of dominant Fourier coefficients such that $x(\tau) \approx  \sum_{k \in \mathcal{U}} \left \langle x \right \rangle_{k}(t) e^{jk\omega_s\tau}$.

\textbf{\textit{Important remarks}}\\
1. Baseband-$abc$ phasor operator $\Upsilon$ is bijective and linear as long as the time-varying phasor's speed is restricted by the low-pass assumption mentioned earlier. On the contrary, no such restriction on the speed is required in the space-phasor-based representation in $dq0$ frame.\\
2. The relationship between the space-phasor $\bar{x}(t)$ and the dynamical sequence components \cite{Stankovic-02-DP} is $\bar{x}(t) = \frac{2}{\sqrt{3}} \sum_{k = -\infty}^{\infty} e^{jk\omega_s t} \left \langle x \right \rangle_{p,k}(t)$. As $k \in \mathcal{U}$ is considered, the generalized averaging-based method leads to an approximated model. On the contrary, the space-phasor-based calculus in $dq0$ frame is an accurate representation as it only depends on a transformation (see Table~\ref{tab:modeling_frameworks}).
\vspace{-10pt}
\begin{table}[h!]
\caption{Properties of balanced-system representation across different phasor calculus frameworks}
\vspace{-5pt}
\label{tab:modeling_frameworks}
\centering
\begin{tabular}{|L{1.9cm}| *{3}{C{1.7cm}|}}
    \hline
        \textbf{Attribute} & \textbf{Baseband-\textit{abc}} & \textbf{Space phasor} & \textbf{Generalized-averaging} \\
    \hline
    Accuracy and speed limitation & accurate under speed limitation  & accurate, no speed limitation & approximate, no speed limitation \\
    \hline
\end{tabular}
\vspace{-5pt}
\end{table}

\noindent 3. For analyzing unbalanced systems and harmonics, generalized averaging gives significant computational advantage compared to $dq0$ frame models.\\
4. For balanced systems, which is the focus of this paper, we consider the space-phasor-based calculus in $d$-$q$ frame as our framework of choice, which we will refer to as the SPC-based approach from now on. For notational convenience, we will use $\bar{x}_{bb} = x_{dq} = x_d + jx_q$ and drop the time variable $t$.

\section{Modeling in SPC and QPC Frameworks}\label{sec:Modeling}
This section briefly discusses the detailed mathematical representation used for modeling a generic power system with SGs, GFLCs, transformers, transmission lines, and loads.  
\vspace{-5pt}
\subsection{Modeling of transmission network and loads}\label{sec:TrNw_LoadMdl}
The transmission network in the SPC-based representation uses a lumped $\pi$-section model as shown in Figs \ref{fig:gen_ref_frame} and \ref{fig:circuit_gflc} consisting of the following KCL and KVL algebraic equations.\\ \vspace{-5pt}
\begin{equation}\label{eq:KCL_VL}
  \mathbf{i_{NDQ}} = \mathbf{CCI}\times \left [ \mathbf{i_{DQ}}^T~\mathbf{i_{lDQ}}^T \right ]^T;~\mathbf{v_{lDQ}} = \mathbf{CCU}\times\mathbf{v_{NDQ}}  
\end{equation}
The model uses a synchronously rotating $D$-$Q$ frame. Assuming the network has $n$ nodes, $l$ series $R$-$L$ branches excluding $m$ SG/IBR transformers, $\mathbf{i_{NDQ}} \in \mathbb{C}^n$ is the vector of net injected currents in each node coming from shunt capacitance and any load that may be present, $\mathbf{i_{lDQ}} \in \mathbb{C}^{l}$ and $\mathbf{i_{DQ}} \in \mathbb{C}^{m}$ are the vectors of currents flowing through each series $R-L$ branch and SG/IBR transformer, respectively, $\mathbf{v_{lDQ}}\in \mathbb{C}^l$ is the vector of voltage drops across series $R$-$L$ branches, $\mathbf{v_{NDQ}} \in \mathbb{C}^n$ is the node voltage vector, and $\mathbf{CCI} \in \mathbb{R}^{n\times (l+m)}$ and $\mathbf{CCU} \in \mathbb{R}^{l\times n}$ are the incidence matrix and nodal connectivity matrix, respectively, from circuit theory.

Constant impedance loads are represented using dynamic models of parallel $R_L$-$L_L$-$C_L$ elements at respective load buses. The following differential equations describe the transmission line and load model, where $\omega_s$ is the synchronous speed in electrical rad/s and the remaining quantities are in per unit (p.u.).
\begin{equation} 
    \begin{array}{l}
         \dot i_{lDQ} = \frac{\omega_{s}}{L_{l}}[ v_{lDQ} - j\omega^* L_{l} i_{lDQ} - R_{l} i_{lDQ}] \\
         \dot v_{NDQ} = \frac{\omega_{s}}{\left(\frac{C_{l}}{2} + C_{L}\right)}\left[i_{NDQ} - i_{LDQ} -\frac{v_{NDQ}}{R_{L}} - j \omega^* (\frac{C_{l}}{2} + C_{L}) v_{NDQ}\right] \\
         \dot i_{LDQ} = \frac{\omega_{s}}{L_{L}}[ v_{NDQ} - j\omega^* L_{L} i_{LDQ}] \\
    \end{array}
\end{equation}
Here, $\omega^* = 1.0$ p.u. and $i_{lDQ}$, $i_{NDQ}$, $v_{lDQ}$, and $v_{NDQ}$ are the elements of corresponding vectors in \eqref{eq:KCL_VL}.

The QPC-based model adopts a transmission network which is represented using admittance matrix-based algebraic equations. Standard current injection framework in the synchronously rotating $D$-$Q$ reference frame is utilized to solve for the bus voltages \cite{pai}. The static loads with constant impedance characteristics are assumed in this model. 

\vspace{-5pt}
\subsection{Modeling of SG including stator transients}\label{sec:SG_Transients}
\begin{figure}[h!]
    \centerline{
\includegraphics[width=0.395\textwidth]{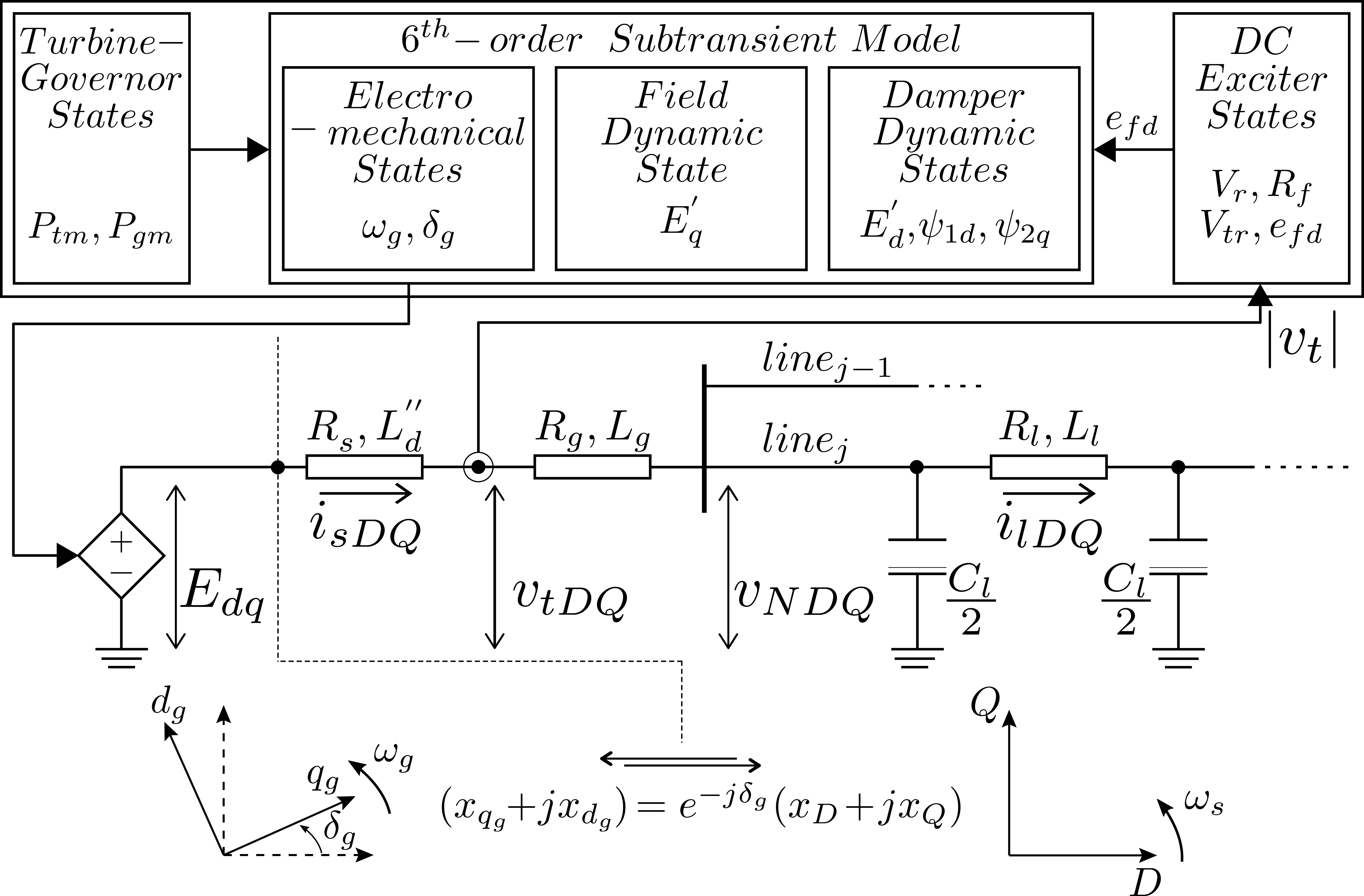}}
    \vspace{-8pt}
    \caption{Interconnection between SG model and transmission network.}
    \label{fig:gen_ref_frame}
    \vspace{-10pt}
\end{figure}

The QPC model of SG considers a $6$th-order subtransient model in its $q_{g}$-$d_{g}$ reference frame rotating at corresponding rotor speed $\omega_g$ along with turbine, governor, and exciter dynamics as shown inside a box in Fig.~\ref{fig:gen_ref_frame}. The model assumes a leading $d$-axis per IEEE convention, neglects stator transients and subtransient saliency (i.e., $L_d^{''} \approx L_q^{''}$), and has  $\omega_{g}$, $\delta_{g}$, $E_{d}^{'}$, $E_{q}^{'}$, $\psi_{1d}$, $\psi_{2q}$ as dynamic states, see pp. $99$ of \cite{chaudhuri2014multi}. 

\textit{The SPC model of SG considers stator transients, whose interface with the dynamic model of the transmission systems has hardly been discussed in classic textbooks like \cite{pai} and \cite{kundur}.} Both of the multi-time-scale models in pp. $91$ of \cite{pai} and in pp. $86$ of \cite{kundur} use stator fluxes $\psi_d$ and $\psi_q$ as state variables, which does not lend itself easily to a current injection framework.

 Therefore, we consider stator currents $i_{sqd}$ ($= i_{sq} + ji_{sd}$) as state variables, combine the $R_g$-$L_g$ dynamics of the transformer with the $R_s$-$L_d^{''}$ dynamics of armature, and pose the Thevenin's equivalent voltage source behind the combined impedance as $E_{qd}$:

\vspace{-10pt}
\begin{equation}
\begin{array}{cc}
     E_{q} = E_{q}^{''} \omega_{g} + \frac{\dot{E}_{d}^{''}}{\omega_{s}} \hspace{30pt}
     E_{d} = E_{d}^{''} \omega_{g} - \frac{\dot{E}_{q}^{''}}{\omega_{s}}
\end{array}
\end{equation}
where, $E_{q}^{''}$ and $E_{d}^{''}$ are as in pp. $100$ of \cite{chaudhuri2014multi}.
As depicted in Fig.~\ref{fig:gen_ref_frame}, using the relationship $x_{DQ} = e^{j\delta_g} x_{q_g d_g}$, we transform all variables to the synchronously rotating $D$-$Q$ frame and in turn write the stator differential equation as 
\begin{equation} \label{eqn:gen_dyn_eqn}
    \begin{array}{l}
         \hspace{-10pt}\dot i_{sDQ} = \frac{\omega_{s}}{(L_{g}+L_{d}^{''})}[ E_{DQ} - v_{NDQ} -j \omega^*(L_{g}+L_{d}^{''}) i_{sDQ} - (R_{g}+R_{s}) i_{sDQ}].
    \end{array}
\end{equation}
\vspace{-15pt}
\vspace{-10pt}
\subsection{Modeling of GFLC and its controls}\label{sec:gflc_modeling}
\begin{figure}[h]
    \centerline{
\includegraphics[width=0.38\textwidth]{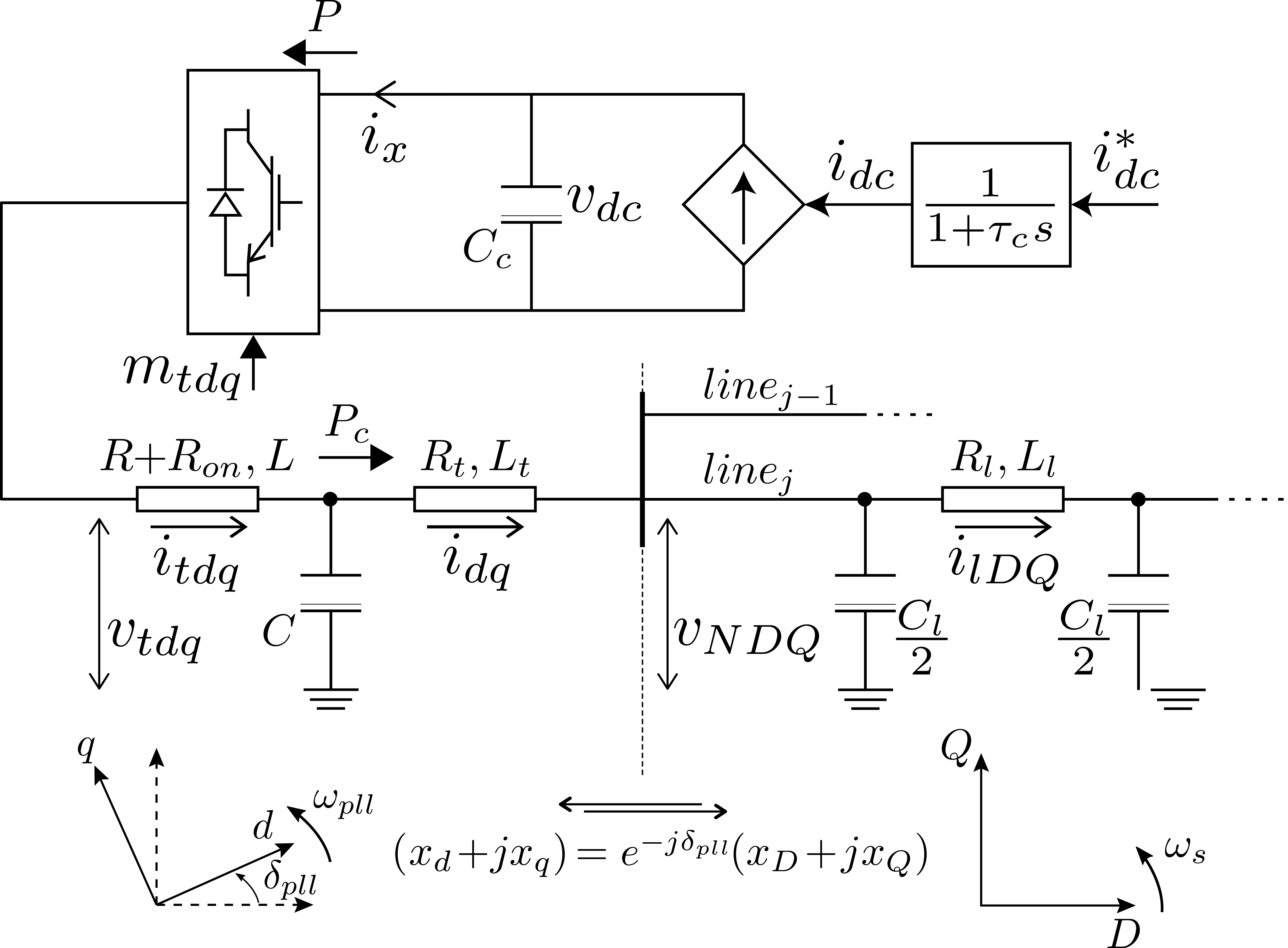}}
    \vspace{-8pt}
    \caption{Circuit model of GFLC [parameters: $\tau_{c}$ = $0.05s$, $C_{c}$ = $1.7370~pu$, $k_{dc}$ = $1080~pu$, $R$ = $0.0033~pu$, $R_{on}$ = $0.0023~pu$, $L$ = $0.2454~pu$, $R_{t}$ = $0~pu$, $L_{t}$ = $0.1500~pu$, $S_{base}$ = $100MVA$, $V_{dc,base}$ = $48.9873kV$, $V_{ac,base}$ = $20kV$].}
    \label{fig:circuit_gflc}
    \vspace{-15pt}
\end{figure}

A typical circuit diagram of a GFLC is illustrated in Fig.~\ref{fig:circuit_gflc}. As shown in the figure, the dc-side of the converter is represented by the dc-link capacitance $C_{c}$ and a current source with a delay $\tau_{c}$, which is the functional model of a renewable energy source. It is noted that converter ac terminal power $P_t = P$ assuming converter losses are added in the form of resistance $R_{on}$ to the filter resistance.

\begin{figure}[h!]
    \centerline{
    \includegraphics[width=0.36\textwidth]{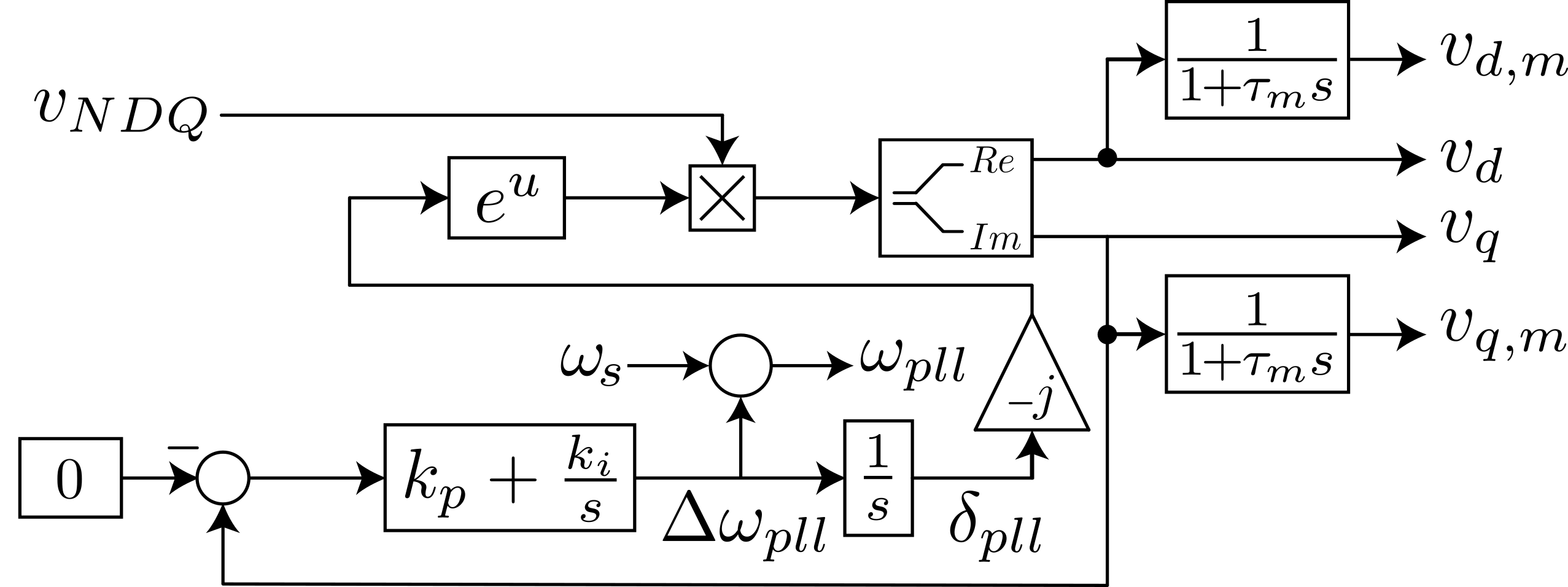}}
    \vspace{-8pt}
    \caption{Phase-locked loop (PLL) [parameters: $k_{p}$ = $101$, $k_{i}$ = $2562$ for $20$ Hz bandwidth and $k_{p}$ = $76$, $k_{i}$ = $1455$ for $15$ Hz bandwidth \cite{pll_impact}, $\tau_{m}$ = $1ms$].}
    \label{fig:pll_control}
    \vspace{-6pt}
\end{figure}

The ac-side quantities of the converter are modeled and controlled in SPC framework, which employs a rotating $d$-$q$ frame generated by the phased-locked loop (PLL), see Fig.~\ref{fig:pll_control}. The ac-side filter, which connects the converter's ac terminal to the grid through a transformer typically consists of $R$-$L$-$C$ elements. However, we neglect the capacitance $C$, which is small and is used to filter out switching harmonics. Thus, the series impedance of the $R$-$L$ filter is combined with series resistance and leakage reactance, $R_{t}$ and $L_{t}$ of the transformer when modeling and that leads to $i_{tdq}$ = $i_{dq}$. It is important to emphasize that the KCL equations representing the dynamics of this circuit segment is in current injection form where injected current $i_{dq}$ must be transformed into $D$-$Q$ frame before using it as an input to the transmission line model. Similarly, the voltage $v_{NDQ}$ must be transformed to $d$-$q$ frame in the KVL equations as shown in \eqref{eqn:con_dyn_eqn}.
\begin{equation} \label{eqn:con_dyn_eqn}
    \begin{array}{l}
         \hspace{-4pt}\dot i_{dq} = \frac{\omega_{s}}{(L + L_{t})}[ v_{tdq} - v_{Ndq} - j\omega_{pll} (L + L_{t}) i_{tdq} - (R_{on} + R + R_{t}) i_{dq}] \\
    \end{array}
\end{equation}

\begin{figure}[h!]
    \centerline{
    \includegraphics[width=0.42\textwidth]{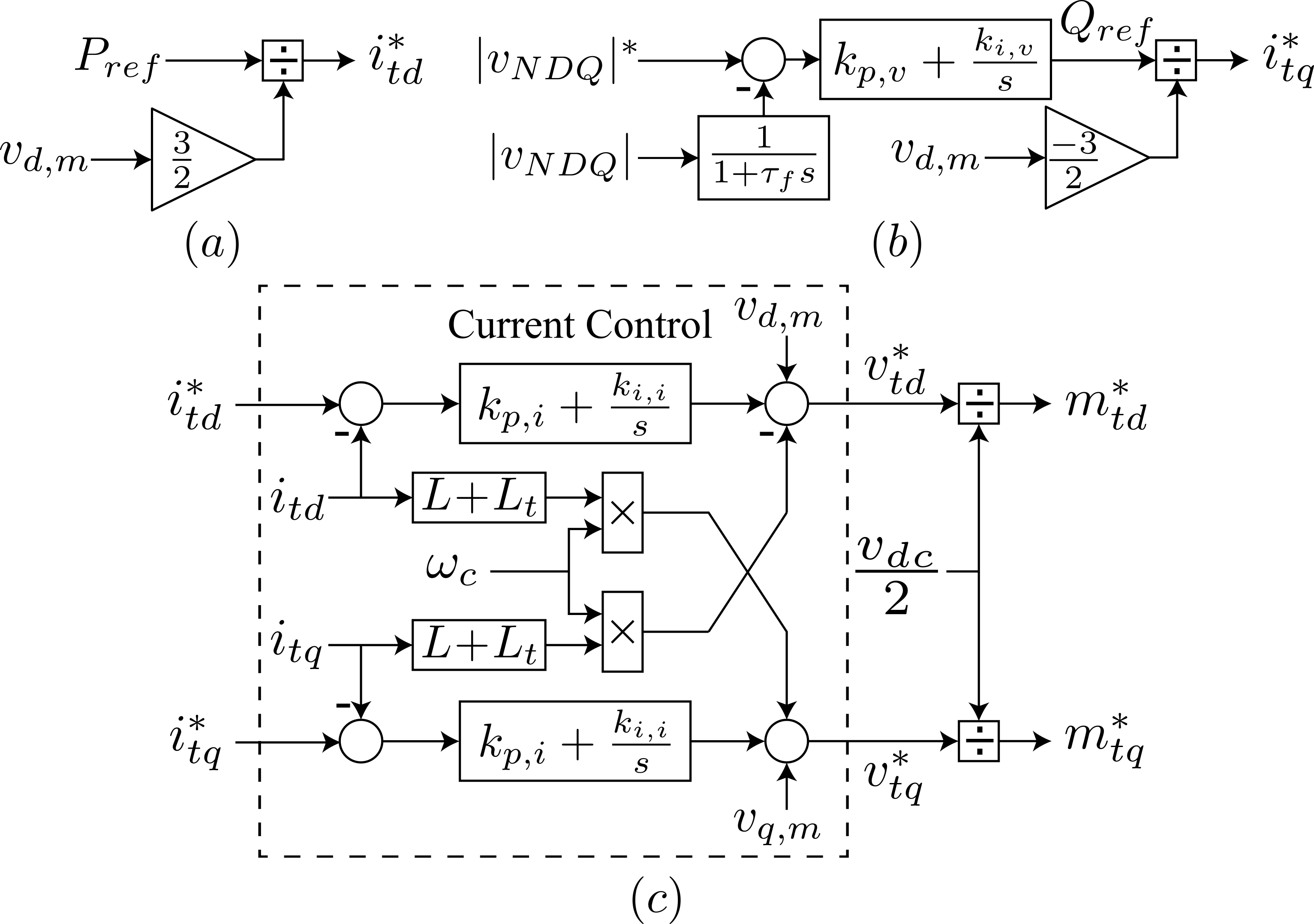}}
    \vspace{-8pt}
    \caption{(a) Active power control, (b) outer voltage control, and (c) inner current control. [parameters: $P_{ref}$ = $7.00pu$, $\tau_{f}$ = $50ms$, $S_{base}$ = $100pu$, $V_{ac,base}$ = $20kV$].}
    \vspace{-17pt}
    \label{fig:gflc_control}
\end{figure}

Figure \ref{fig:pll_control} shows the block diagram of the PLL model. The PLL facilitates synchronism by aligning $d$ axis of the reference frame along that of the voltage vector $v_{NDQ}$ and maintains a zero $q$-axis component. It estimates the phase angle $\delta_{pll}$ of the voltage $v_{NDQ}$ at the point of common coupling (PCC) for coordinate transformation utilizing the angular frequency $\omega_{pll}$. Variables $v_d$ and $v_q$ shown in Fig.~\ref{fig:pll_control} are used for ac-dc power balance equations, whereas filtered versions $v_{d,m}$ and $v_{q,m}$ are used in the control loops shown in Fig.~\ref{fig:gflc_control}.

The standard inner current control loops as shown in Fig.~\ref{fig:gflc_control}(c) are employed in regulating current through the series $R-L$ filter branch and that of the transformer. Current references $i_{td}^*$ and $i_{tq}^*$ are generated by the active power control and outer voltage control loop, respectively, see Figs~\ref{fig:gflc_control}(a) and (b). As illustrated in the figure, the reactive power reference is generated by a PI controller which regulates the voltage magnitude $|v_{NDQ}|$ at the PCC.

Table~\ref{tab:modeling_components} summarizes how various components of the system are represented in different modeling frameworks.
\vspace{-9pt}
\begin{table}[h!]
\caption{Summary of component modeling in different frameworks}
\label{tab:modeling_components}
\vspace{-5pt}
\centering
\begin{tabular}{|L{1.5cm}| *{2}{C{2.6cm}|}}
    \hline
    \multirow{2}{*}{\textbf{Component}} &
        \multicolumn{2}{c|}{\textbf{Modeling framework}} \\
        \cline{2-3}
        & \textbf{QPC} & \textbf{SPC}\\
    \hline
    SG stator transients & neglected & dynamically modeled \newline[Section \ref{sec:SG_Transients}] \\
    \hline
    Transmission network & \hspace{15pt}Y-bus \newline[Section \ref{sec:TrNw_LoadMdl}] & dynamically modeled [Section \ref{sec:TrNw_LoadMdl}]\\
    \hline
    Loads & static load model \newline[Section \ref{sec:TrNw_LoadMdl}] & dynamic load model \newline[Section \ref{sec:TrNw_LoadMdl}]\\
    \hline
    GFLC & \multicolumn{2}{c|}{identically modeled [Section \ref{sec:gflc_modeling}]} \\
    \hline
\end{tabular}
\vspace{-9pt}
\end{table}

\vspace{-4pt}
\section{Proposed Approach for MIMO Modeling Adequacy Study in Small-Signal Sense}\label{sec:MdlAdeq}
\vspace{2pt}
It is a well-established fact that QPC-based models are an accurate enough representation to perform dynamic stability studies of grids involving traditional SGs. However, with the IBR integration to the grid, this aspect must be revisited as critical and dominant modal frequencies go beyond electro-mechanical range. 
\textit{If QPC-based models can be validated as an adequate representation of a given system for studying certain phenomena, this would lead to substantial savings in time and computational resources when conducting exhaustive simulations.}
Thus motivated, we propose a systematic approach to determine the modeling adequacy of QPC-based models in small-signal sense, comparing with SPC-based models which in this case is considered a more detailed representation of the system. In fact, the proposed approach can be applied to any generic dynamical system to investigate adequacy of MIMO models.

For convenience, we refer the \textit{linearized} versions of the detailed model (e.g., SPC-based model) as $G$ and the approximate model under consideration (e.g., QPC-based model) as $G_r$. The steps in analyzing the modeling adequacy are summarized next, which are followed by a detailed discussion. 
The approach is further elaborated by findings presented in Section \ref{sec:Results6bus} along with the case studies.

Consider the linearized MIMO model $G$ in the descriptor state-space or transfer function form

\begin{equation}\label{eq:state_G}{
\vspace{-5pt}
\small
\begin{array}{*{20}c}
G:\left\{ {\begin{array}{*{20}c}
	{E \dot x(t) = A x(t) + B u(t)}  \\
	{y(t) = C x(t) + D u(t)}  \\
	\end{array}} \right.\\
or,\\
{G(s) = C(sE - A)^{ - 1} B + D } \in \mathbb{C}^{p\times m}
\end{array}}
\end{equation}
\normalsize
where, ${A, E\in{\mathbb{R}}^{n\times n}}$, ${B\in{\mathbb{R}}^{n\times m}}$, ${C\in{\mathbb{R}}^{p\times n}}$, ${D\in{\mathbb{R}}^{p\times m}}$, ${x(t)\in{\mathbb{R}}^{n}}$, ${u(t)\in {\mathbb{R}}^{m}}$, ${y(t)\in {\mathbb{R}}^{p}}$. Similarly, the approximate model $G_r$ can be expressed as
\begin{equation}\label{eq:state_Gr}{
\vspace{-5pt}
\small
\begin{array}{*{20}c}
G_r:\left\{ {\begin{array}{*{20}c}
		{E_r\dot x(t) = A_r x(t) + B_r u(t)}  \\
		{y(t) = C_r x(t) + D_r u(t)}  \\
		\end{array}} \right.\\
	or,\\
	{G_r(s) = C_r(sE_r - A_r)^{ - 1} B_r + D_r } \in \mathbb{C}^{p\times m}
  \end{array}}
\end{equation}
\normalsize
having much smaller dimension ${r<<n}$ with ${A_r,E_r\in{\mathbb{R}}^{r\times r}}$, ${B_r\in{\mathbb{R}}^{r\times m}}$, ${C_r\in{\mathbb{R}}^{p\times r}}$, ${D_r\in{\mathbb{R}}^{p\times m}}$, ${x(t)\in{\mathbb{R}}^{r}}$, ${u(t)\in {\mathbb{R}}^{m}}$, ${y(t)\in {\mathbb{R}}^{p}}$.

\vspace{5pt}
\setlist[enumerate,1]{leftmargin=35pt}
\setlist[enumerate,2]{leftmargin=10pt}
\begin{enumerate}
    \item[Step 1:] Consider the frequency range ($0-f^*$) up to which $G$'s accuracy is guaranteed.
    \item[Step 2:] Within the range obtained from Step $1$,
    \begin{enumerate}
        \item consider the pole maps for both the models and find poorly-damped poles (critical modes) considering a damping ratio threshold $\zeta^*$.
        \item check singular value plots to determine `dominance' of the poles.
        \item characterize the dominant critical modes using participation factor analysis.
    \end{enumerate}
    \item[Step 3:] If $G_r$ has characteristically different dominant and critical modes from $G$, then $G_r$ is \underline{not} adequate. Otherwise, check if dominant and critical poles of $G$ are (closely) represented in $G_r$ as well. 
    \item[Step 4:] Finally, assess the \textit{weighted} maximum singular value error plots to gain insight into the relative accuracy of $G_r$ w.r.t. $G$.
    \item[Step 5:] (a) Time-domain, and (b) sensitivity analysis should be performed for cross-validation.
\end{enumerate}

\vspace{5pt}
\subsubsection{Step 1}
As mentioned earlier, $G$ is considered to be the benchmark, which is a more detailed representation of the system compared to $G_r$. However, even this model has a finite bandwidth beyond which the accuracy starts to degrade. Thus, in this step, we determine the frequency $f^*$ up to which this model can be trusted. This may be easy for a very simple system, but as the order of the system increases, it becomes difficult. In such a situation, component-wise analysis can help determine which component's properties may restrict the overall system's bandwidth as prior knowledge on the frequency range over which each key component of the power system is accurate have been well-studied in the literature. For example, in the SPC-based model considered in our study, the lumped-$\pi$-section transmission line model is the determining factor, which limits the overall bandwidth to the synchronous frequency. Thus, modeling adequacy studies should not be performed for frequencies over $f^* = 60$ Hz.

\subsubsection{Step 2}
Once the range is decided, it is important to identify what are the most dominant and critical modes in this range. To accomplish that, pole maps of each model need to be examined. Typically, poorly-damped modes are considered as \textit{critical} whereas modes with large gains in the form of maximum singular values $\sigma_{max}(j\omega)$ are considered as \textit{dominant}, e.g., the maximum singular values of $G$ are defined as $\sigma_{max}(G(j\omega)) = \underset{d(j\omega)\neq0}{max} \frac{\left \| G(j\omega)d(j\omega) \right \|_2}{\left \| d(j\omega) \right \|_2}$. 
Participation factor analysis \cite{verghese_participationf} should be used to characterize the modes by associating them with states. 

\subsubsection{Step 3}
In the modeling adequacy studies, it is important to 
guarantee that $G_r$ does not exhibit any false dynamics. 
If the approximate model $G_r$ exhibits characteristically different dominant critical modes from $G$, then $G_{r}$ is ruled out. If such inconsistencies are not observed, we compare the pole maps to make sure that the modes of interest in $G$ are also (closely) represented in $G_r$.

\subsubsection{Step 4} 
Finally, maximum singular value error ($\bar \sigma_{e}(j\omega)$)  between \textit{input weighted} models $G(s)\mathcal{W}(s)$ and $G_r(s)\mathcal{W}(s)$ are assessed, where 
\begin{equation}
   \bar \sigma_{e} (j\omega) = | \sigma_{max} (G(j\omega)\mathcal{W}(j\omega)) - \sigma_{max} (G_r(j\omega)\mathcal{W}(j\omega)) |
\end{equation}
and $\mathcal{W}(s) \in \mathbb{C}^{m\times m}$ is a given \textit{input shaping filter}. We propose the filter to be of the simple form $\mathcal{W}(s) = w(s) I$, where $w(s) \in \mathbb{C}$ and $I \in \mathbb{R}^{m \times m}$ is the identity matrix. We choose $\mathcal{W}(s)$ to be a low pass or band-pass filter, which facilitates the weighting over a certain range of frequency. The cut-off frequencies are decided based on $f^*$ from \textit{Step 1}.
  
We propose two metrics to quantify the modeling error --\\
1. $\sigma_e^* = \underset{\omega}{sup}~\bar{\sigma}_e(j\omega)$ and 
2. $\sigma_e^{crit} = \bar{\sigma}_e(j\omega^{crit})$, where $\omega^{crit}$ corresponds to the frequency of a critical mode. These measures quantify the accuracy of $G_r$ w.r.t. $G$.

\subsubsection{Step 5} (a) It is important to cross-validate the adequacy by comparing time-domain simulation results with EMT simulations, if possible (which is difficult for large-scale systems). (b) Moreover, it is critical to perform sensitivity analysis in frequency domain by varying important system parameters to ensure that modeling adequacy of $G_r$ is valid across a range.

\section{Results and Analysis}\label{sec:Results6bus}
\begin{figure}[h!]
    \centerline{
    \includegraphics[width=0.48\textwidth]{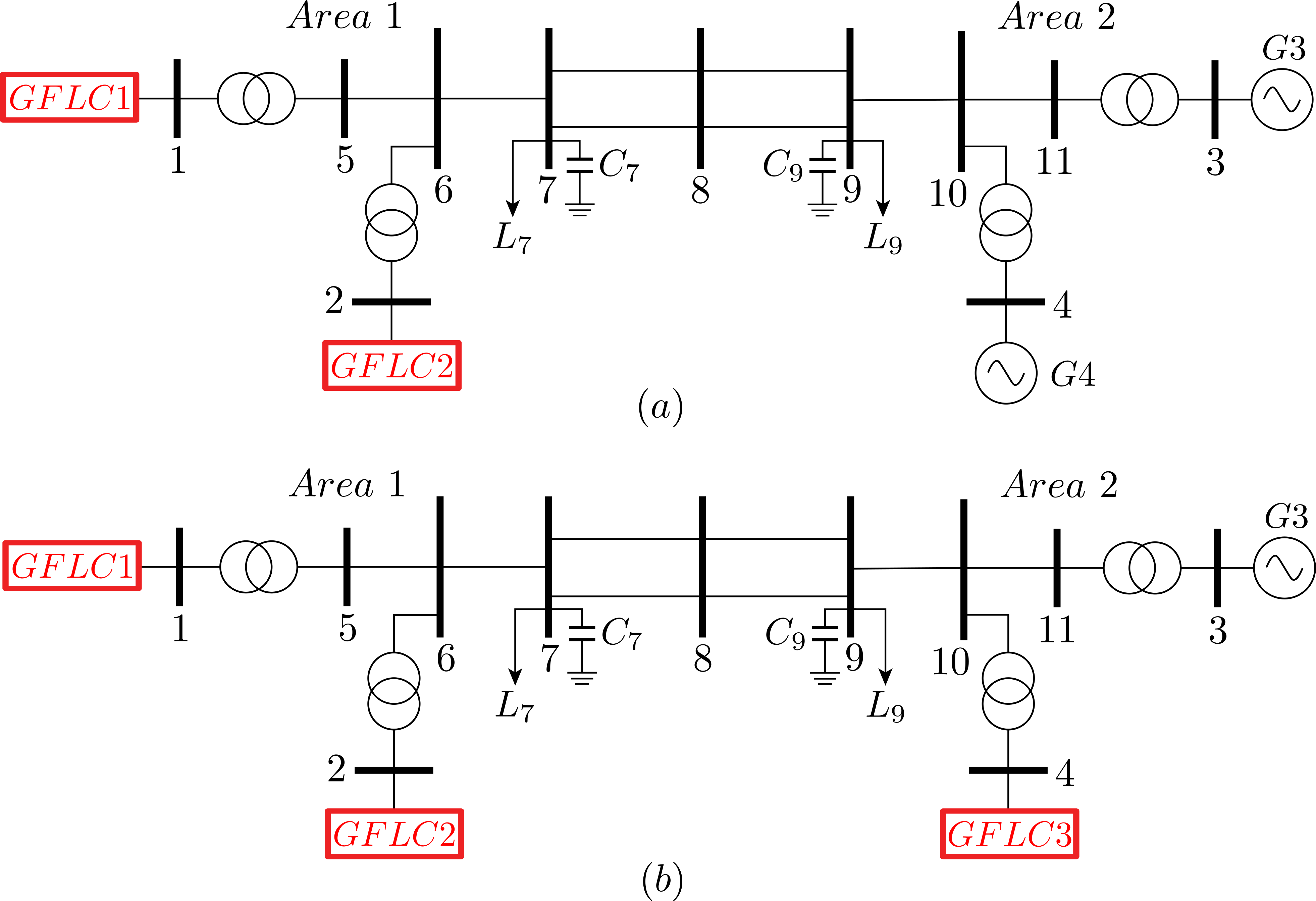}}
    \vspace{-8pt}
    \caption{Modified two-area test system with -- (a) case (1): $50$\% and (b) case (2): $75$\% IBR penetration }
    \label{fig:11bus_sys}
    \vspace{-14pt}
\end{figure}

This section focuses on case studies performed on the IEEE $4$-machine test system for different levels of IBR penetration after replacing some of the SGs with \textit{identical} GFLCs while operating under nominal condition with $400$ MW tie-flow. Our findings are primarily based on two case studies -- Case (1): an analysis on the modified two-area test system with $2$ GFLCs and $2$ SGs (see, Fig.~\ref{fig:11bus_sys}(a)) followed by Case (2): an analysis on the same test system with $3$ GFLCs and $1$ SG (see, Fig.~\ref{fig:11bus_sys}(b)). Both the cases are comprehensively analyzed to validate the proposed approach of evaluating modeling adequacy. To that end, QPC and SPC-based models of the test systems shown in Fig.~\ref{fig:11bus_sys} are implemented in Matlab/Simulink as described in Section~\ref{sec:Modeling}. The step-by-step analysis is elaborated next.

\textit{\textbf{Step 1:}} As mentioned earlier, we consider the SPC model is valid up to $f^* = 60$ Hz due to the lumped $\pi$ model of transmission lines. 

As pointed out in the Section \ref{sec:MdlAdeq}, it is crucial to identify dominant and/or critical modes in a dynamic system. For that purpose, first we perform frequency-domain analysis on MIMO linearized QPC and SPC-based models for both cases under consideration. The inputs of these models are $\Delta \omega_{pll}$ and $Q_{ref}$ of each converter and outputs are $v_q$ and $|v_{NDQ}|$ of respective converters. Corresponding to \eqref{eq:state_G} and \eqref{eq:state_Gr} for case (1): $n = 86$, $r = 48$, $m = 4$, and $p = 4$; and for case (2): $n = 84$, $r = 48$, $m = 6$, and $p = 6$.  The system parameters and some of the controller parameters are specified under the captions of Figs~\ref{fig:circuit_gflc},~\ref{fig:pll_control}, and \ref{fig:gflc_control}. As per \cite{yazdani}, the inner current controller's closed-loop bandwidth can range from $200$-$2000$ rad/s. In the nominal case we have chosen a $300$ rad/s bandwidth leading to the current controller parameters $k_{p,i}$ = $0.0350$ pu, $k_{i,i}$ = $0.1835$ pu (Fig.~\ref{fig:gflc_control}). The nominal voltage controller parameters are $k_{p,v}$ = $5$ pu and $k_{i,v}$ = $0.5$ pu, based on \cite{full_converter_wind}. 

\begin{figure}[h!]
    \centerline{
    \includegraphics[width=0.46\textwidth]{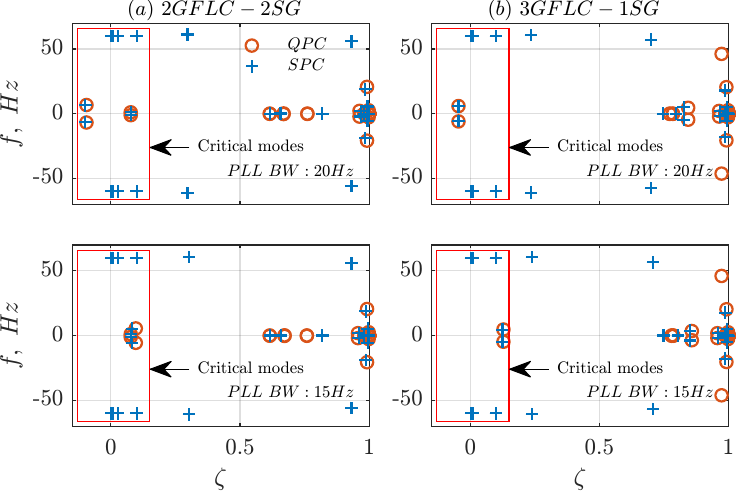}}
    \vspace{-8pt}
    \caption{Pole map of linearized QPC and SPC-based models for (a). case (1) and (b). case (2) considering PLL bandwidths $20$ Hz and $15$ Hz.}
    \label{fig:eigenv_comp1}
    \vspace{-4pt}
\end{figure}

\textit{\textbf{Step 2a:}} Figure \ref{fig:eigenv_comp1} shows the frequency ($f$) against damping ratio ($\zeta$) of eigenvalues for each scenario considering $f^* = 60$ Hz. Assuming $\zeta^* = 0.15$, it is observed in both cases ($1$) and ($2$) that the QPC and SPC-based models exhibit a critical stable mode of $\approx 5$ Hz and unstable mode of $\approx 6$ Hz for PLL bandwidths $15$ Hz and $20$ Hz, respectively. In addition, a mode with $\approx 1$ Hz also appears in this range. Moreover, a few pairs of critical $60$ Hz modes show up exclusively in the SPC model.

\begin{figure}[h!]
    \centerline{
    \includegraphics[width=0.48\textwidth]{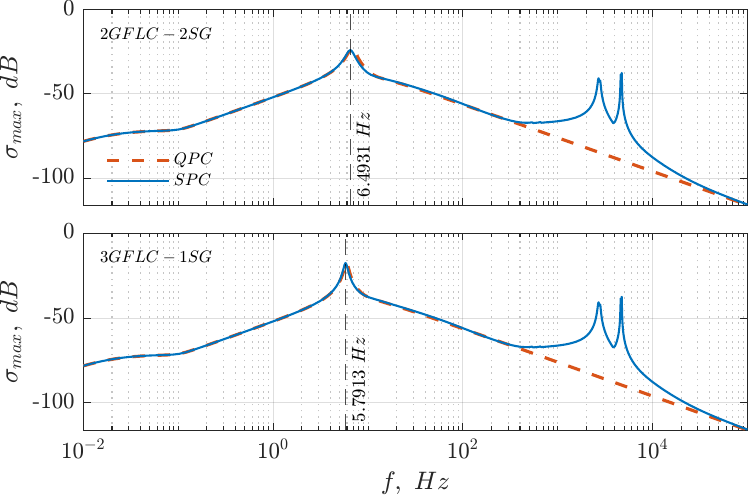}}
    \vspace{-8pt}
    \caption{Comparison of maximum singular values of QPC and SPC models for $20$ Hz PLL bandwidth.}
    \label{fig:sv_plot_20Hz}
\end{figure}

\begin{figure}[h!]
    \centerline{
    \includegraphics[width=0.48\textwidth]{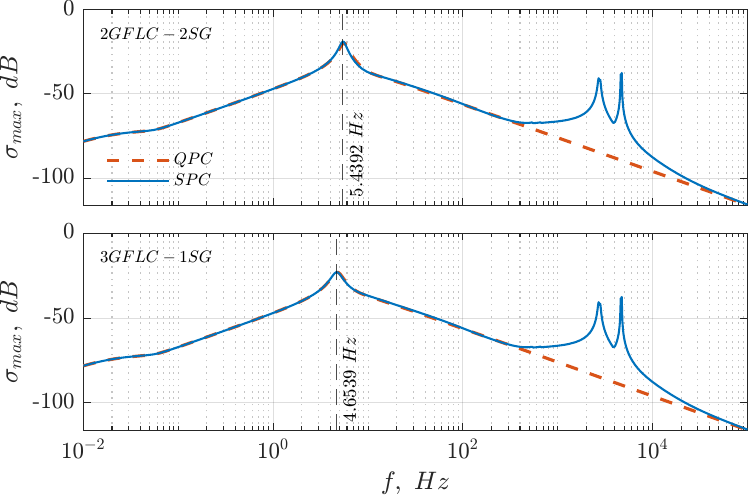}}
    \vspace{-8pt}
    \caption{Comparison of maximum singular values of QPC and SPC models for $15$ Hz PLL bandwidth.}
    \vspace{-0pt}
    \label{fig:sv_plot_15Hz}
\end{figure}

\textbf{\textit{Step 2b:}}
Figures~\ref{fig:sv_plot_20Hz} and \ref{fig:sv_plot_15Hz} show the singular value plots of different MIMO models, which indicate that in none of the cases, any peaks in maximum singular values are observed near $1$ Hz and $60$ Hz frequencies, thereby ruling them out as dominant modes. The peaks observed in the higher frequency range, attributed to network modes, should not be considered, as the reliability of our models in that range is questionable. Sharp peaks can be observed near $5$ Hz and $6$ Hz in Figs \ref{fig:sv_plot_15Hz} and \ref{fig:sv_plot_20Hz}, respectively, which make those modes both critical and dominant. Tables~\ref{tab:freq_unstable_mode_2gflc} and \ref{tab:freq_unstable_mode_3gflc} summarize the frequencies and damping ratios of these modes obtained from models of different frameworks and distinct configurations. Thus, these modes are of greater significance in the progression of our discussion.
\vspace{-5pt}
\vspace{-15pt}
\vspace{-5pt}
\begin{table}[h!]
\scriptsize
\caption{Case (1): SSO mode in different modeling frameworks}
\label{tab:freq_unstable_mode_2gflc}
\vspace{-5pt}
\centering
\begin{tabular}{| l | *{3}{c|} }
    \hline
    \multicolumn{4}{|c|}{\textbf{PLL BW: $15$ Hz}} \\
    \hline
    \textbf{Framework} & \textbf{EMT} & \textbf{SPC} & \textbf{QPC}\\
    \hline
    Approach & Prony & Linearization & Linearization\\
    \hline
    $f, Hz$ & $5.5030$ & $5.4392$ & $5.5645$\\
    \hline
    $\zeta, \%$ & $13.79$ & $8.24$ & $9.64$\\
    \hline
    \multicolumn{4}{|c|}{\textbf{PLL BW: $20$ Hz}} \\
    \hline
    \textbf{Framework} & \textbf{EMT} & \textbf{SPC} & \textbf{QPC}\\
    \hline
    Approach & Prony & Linearization & Linearization\\
    \hline
    $f, Hz$ & $6.5580$ & $6.4931$ & $6.7550$\\
    \hline
    $\zeta, \%$ & $-11.48$ & $-9.93$ & $-9.47$\\
    \hline
\end{tabular}
\vspace{-10pt}
\end{table}

\begin{table}[h!]
\scriptsize
\caption{Case (2): SSO mode in different modeling frameworks}
\label{tab:freq_unstable_mode_3gflc}
\vspace{-5pt}
\centering
\begin{tabular}{| l | *{3}{c|} }
    \hline
    \multicolumn{4}{|c|}{\textbf{PLL BW: $15$ Hz}} \\
    \hline
    \textbf{Framework} & \textbf{EMT} & \textbf{SPC} & \textbf{QPC}\\
    \hline
    Approach & Prony & Linearization & Linearization\\
    \hline
    $f, Hz$ & $4.6150$ & $4.6539$ & $4.7140$\\
    \hline
    $\zeta, \%$ & $19.31$ & $12.46$ & $12.84$\\
    \hline
    \multicolumn{4}{|c|}{\textbf{PLL BW: $20$ Hz}} \\
    \hline
    \textbf{Framework} & \textbf{EMT} & \textbf{SPC} & \textbf{QPC}\\
    \hline
    Approach & Prony & Linearization & Linearization\\
    \hline
    $f, Hz$ & $5.8310$ & $5.7913$ & $5.9220$\\
    \hline
    $\zeta, \%$ & $-5.42$ & $-4.58$ & $-4.57$\\
    \hline
\end{tabular}
\vspace{-10pt}
\end{table}

\begin{figure}[h!]
    \centerline{
    \includegraphics[width=0.5\textwidth]{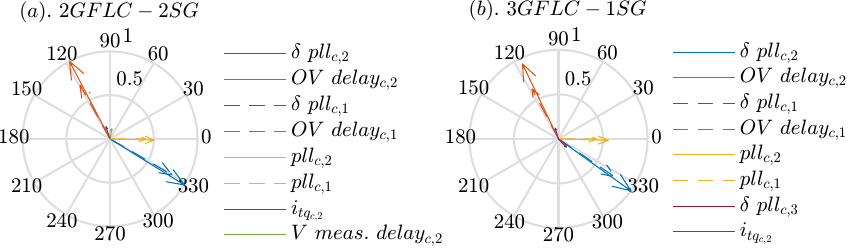}}
    \vspace{-5pt}
    \caption{SPC model: Compass plots of normalized participation factor magnitudes and modeshape angles of the dominant states contributing to the unstable eigenvalue pair for $20$ Hz PLL bandwidth. (a) Case (1), (b) Case (2).}
    \label{fig:mode_shapes_SPC}
    \vspace{-10pt}
\end{figure}

\begin{figure}[h!]
    \centerline{
    \includegraphics[width=0.5\textwidth]{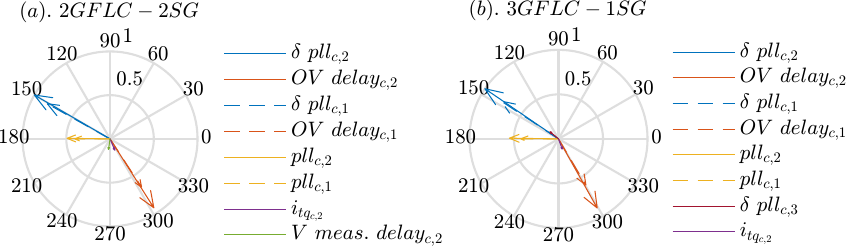}}
    \vspace{-5pt}
    \caption{QPC model: Compass plots of normalized participation factor magnitudes and modeshape angles of the dominant states contributing to the unstable eigenvalue pair for $20$ Hz PLL bandwidth. (a) Case (1), (b) Case (2).}
    \label{fig:mode_shapes_QPC}
    \vspace{-15pt}
\end{figure}

\textbf{\textit{Step 2c:}}
Figures~\ref{fig:mode_shapes_SPC} and \ref{fig:mode_shapes_QPC} show compass plots of normalized participation factor magnitudes and modeshape angles of the dominant states contributing to the unstable mode for $20$ Hz PLL bandwidth, respectively for SPC and QPC-based models. It is evident that the PLL states and states corresponding to outer voltage control loop delays in the feedback signals have high dominance in these eigenvalue pairs for both SPC and QPC models. \textit{Therefore, the dominant and critical mode is characterized as IBR SSO mode.} Although not shown here, it is also found that this fact remains the same for the $15$ Hz PLL bandwidth.   
\vspace{-3pt}
\vspace{-3pt}

\begin{figure}[h!]
    \centerline{
    \includegraphics[width=0.46\textwidth]{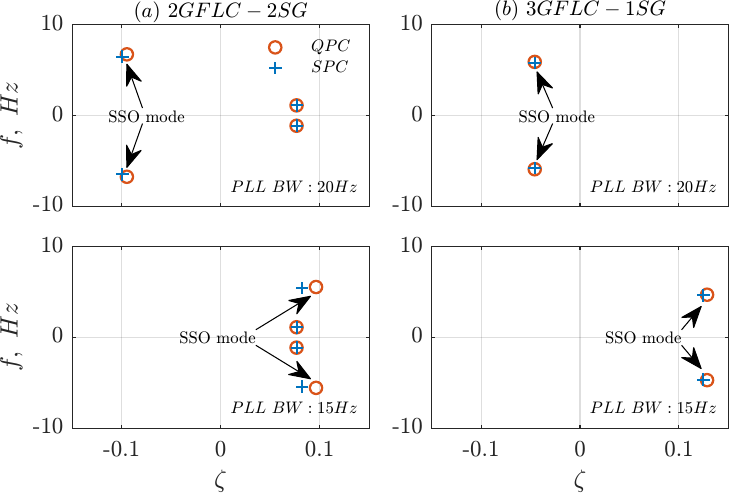}}
    \vspace{-8pt}
    \caption{Zoomed pole map of linearized QPC and SPC-based models for (a). case (1) and (b). case (2) considering PLL bandwidths $20$ Hz and $15$ Hz capturing critical modes of interest.}
    \label{fig:eigenv_comp2}
    \vspace{-15pt}
\end{figure}

\textbf{\textit{Step 3:}} The dominant and critical IBR SSO modes are closely represented in the QPC-based models as observed from Tables~\ref{tab:freq_unstable_mode_2gflc} and \ref{tab:freq_unstable_mode_3gflc} (also see zoomed views in Fig.~\ref{fig:eigenv_comp2}), which leads us to the next step of the adequacy analysis.

\begin{figure}[h!]
    \centerline{
    \includegraphics[width=0.46\textwidth]{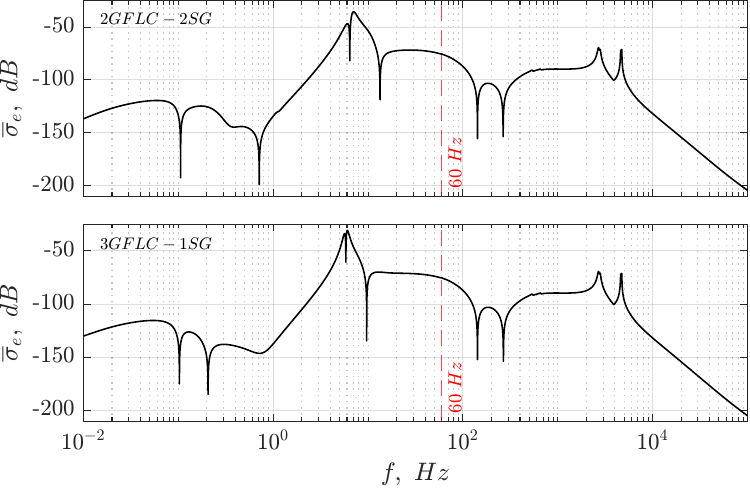}}
    \vspace{-8pt}
    \caption{Maximum singular value errors for $20$ Hz PLL bandwidth.}
    \label{fig:svmax_error_20}
    \vspace{-5pt}
\end{figure}

\begin{figure}[h!]
    \centerline{
    \includegraphics[width=0.46\textwidth]{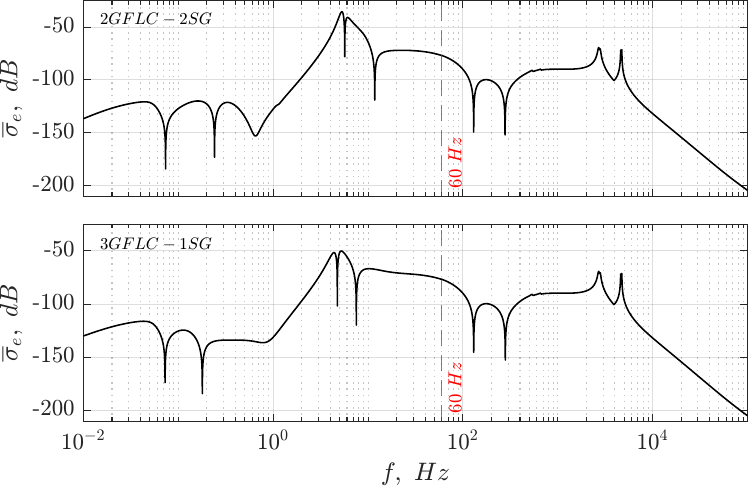}}
    \vspace{-8pt}
    \caption{Maximum singular value errors for $15$ Hz PLL bandwidth.}
    \label{fig:svmax_error_15}
    \vspace{-15pt}
\end{figure}

\textbf{\textit{Step 4:}}
In this step we present metrics to quantify the modeling adequacy of the MIMO QPC model w.r.t. the SPC model. We consider $w(s) = \frac{1}{1 + s\tau}$ with a corner frequency of $100$ Hz for the input shaping filter, which ensures a unity gain up to about $60$ Hz. Figures ~\ref{fig:svmax_error_20} and ~\ref{fig:svmax_error_15} show the maximum singular value error ($\bar{\sigma}_{e}$) of QPC-based models w.r.t. SPC models versus frequency for PLL bandwidths of $20$ Hz and $15$ Hz, respectively. It is clear that the largest maximum error $\sigma_e^*$  between the two models for various scenarios and $\sigma_e^{crit}$ at SSO modal frequencies of interest are very small as indicated in Tables~\ref{tab:error_norms1} and \ref{tab:error_norms2}. 
Thus, we can conclude that the QPC-based model is adequate to capture SSOs accurately w.r.t. the SPC-based model in these scenarios within the frequency range $0$-$60$ Hz. 

\vspace{-12pt}
\begin{table}[h!]
\scriptsize
\caption{Case (1): Weighted maximum singular value errors}
\label{tab:error_norms1}
\vspace{-5pt}
\centering
\begin{tabular}{| l | *{3}{c|} }
    \hline
    \multicolumn{4}{|c|}{\textbf{PLL BW: $15$ Hz}}\\
    \hline
    \multirow{2}{*}{$f, Hz$} & 
        $f_{SSO,SPC}$ & $f_{SSO,QPC}$ & $f_{\sigma_{e,max}}$\\
        \cline{2-4}
        & $5.4392$ & $5.5645$ & $5.3048$ \\
    \hline
    $\overline \sigma_{e},\ dB $ & $-36.8859$ & $-42.6454$ & $-35.3609$ \\
    \hline
    \multicolumn{4}{|c|}{\textbf{PLL BW: $20$ Hz}}\\
    \hline
    \multirow{2}{*}{$f, Hz$} & 
        $f_{SSO,SPC}$ & $f_{SSO,QPC}$ & $f_{\sigma_{e,max}}$\\
        \cline{2-4}
        & $6.4931$ & $6.7550$ & $7.0451$ \\
    \hline
    $\overline \sigma_{e},\ dB $ & $-52.7573$ & $-38.1476$ & $-35.4485$ \\
    \hline
\end{tabular}
\vspace{-18pt}
\end{table}

\begin{table}[h!]
\scriptsize
\caption{Case (2): Weighted maximum singular value errors}
\label{tab:error_norms2}
\vspace{-5pt}
\centering
\begin{tabular}{| l | *{3}{c|} }
    \hline
    \multicolumn{4}{|c|}{\textbf{PLL BW: $15$ Hz}}\\
    \hline
    \multirow{2}{*}{$f, Hz$} & 
        $f_{SSO,SPC}$ & $f_{SSO,QPC}$ & $f_{\sigma_{e,max}}$\\
        \cline{2-4}
        & $4.6539$ & $4.7140$ & $5.2200$ \\
    \hline
    $\overline \sigma_{e},\ dB $ & $-59.0848$ & $-67.7766$ & $-50.1939$ \\
    \hline
    \multicolumn{4}{|c|}{\textbf{PLL BW: $20$ Hz}}\\
    \hline
    \multirow{2}{*}{$f, Hz$} & 
        $f_{SSO,SPC}$ & $f_{SSO,QPC}$ & $f_{\sigma_{e,max}}$\\
        \cline{2-4}
        & $5.7913$ & $5.9220$ & $6.0545$ \\
    \hline
    $\overline \sigma_{e},\ dB $ & $-40.4508$ & $-34.9837$ & $-30.7342$ \\
    \hline
\end{tabular}
\vspace{-10pt}
\end{table}

\begin{figure}[h!]
    \centerline{
    \includegraphics[width=0.48\textwidth]{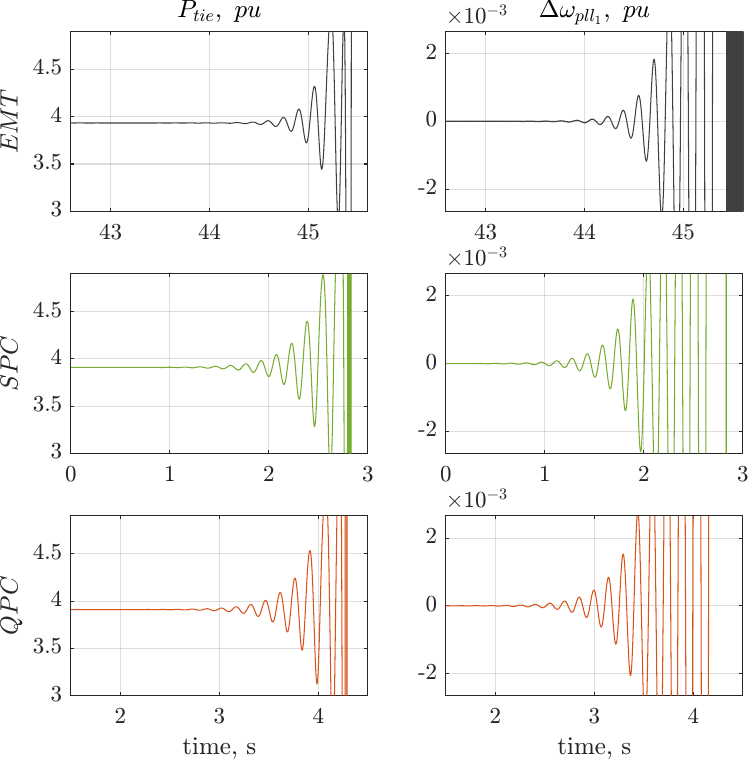}}
    \vspace{-8pt}
    \caption{Case (1): Comparison of dynamic behavior of the $2$GFLC-$2$SG system modeled in different frameworks for $20$ Hz PLL bandwidth.}
    \vspace{-10pt}
    \label{fig:20Hz_2gflc}
\end{figure}

\begin{figure}[h!]
    \centerline{
    \includegraphics[width=0.48\textwidth]{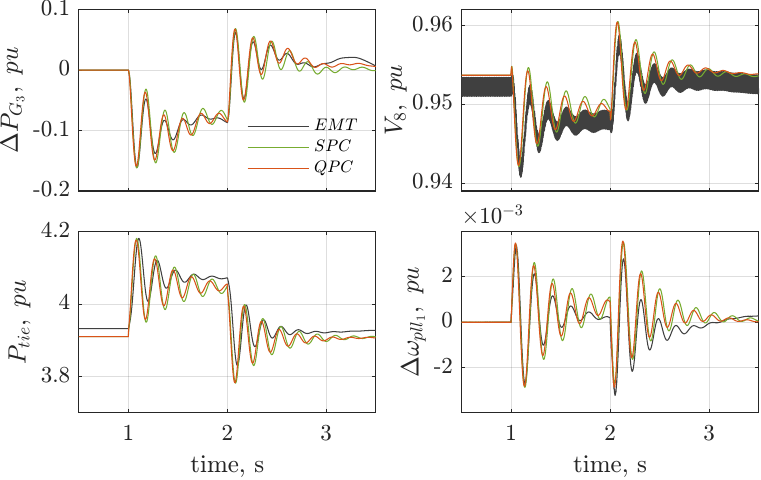}}
    \vspace{-8pt}
    \caption{Case (1): Comparison of dynamic behavior of the $2$GFLC-$2$SG system modeled in different frameworks following a $1$s pulse change in active power reference of the GFLC$1$ for $15$ Hz PLL bandwidth.}
    \vspace{-10pt}
    \label{fig:15Hz_2gflc_2}
\end{figure}

\begin{figure}[h!]
    \centerline{
    \includegraphics[width=0.48\textwidth]{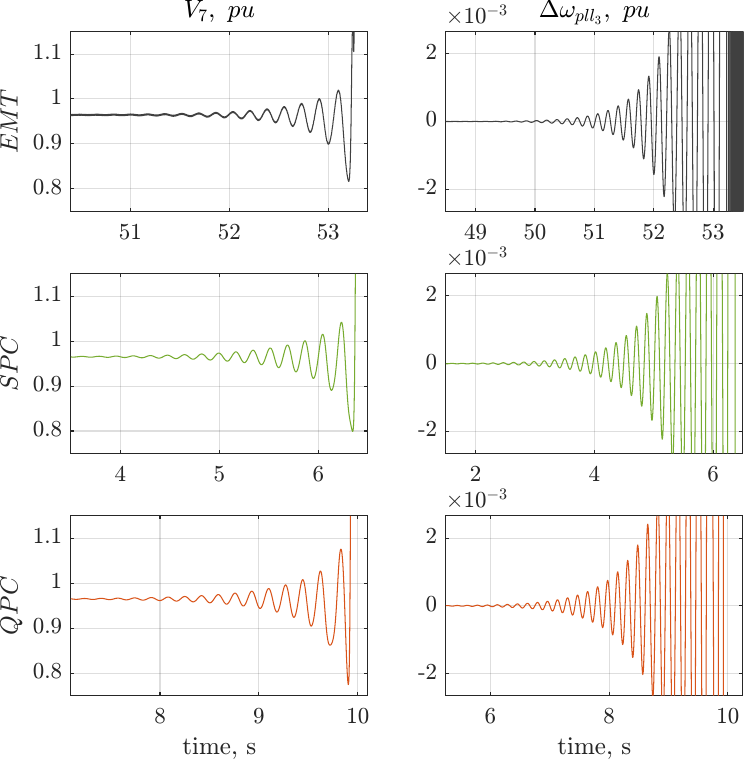}}
    \vspace{-8pt}
    \caption{Case (2): Comparison of dynamic behavior of the $3$GFLC-$1$SG system modeled in different frameworks for $20$ Hz PLL bandwidth.}
    \vspace{-5pt}
    \label{fig:20Hz_3gflc}
\end{figure}

\begin{figure}[h!]
    \centerline{
    \includegraphics[width=0.48\textwidth]{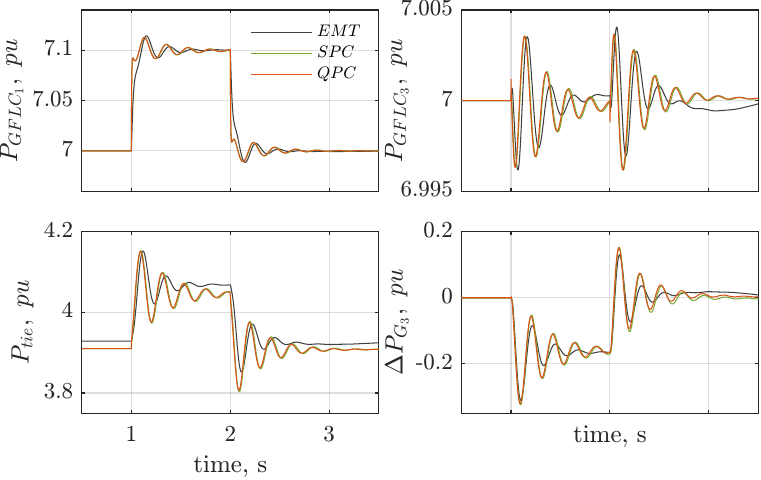}}
    \vspace{-8pt}
    \caption{Case (2): Comparison of dynamic behavior of the $3$GFLC-$1$SG system modeled in different frameworks following a $1$s pulse change in active power reference of the GFLC$1$ for $15$ Hz PLL bandwidth.}
    \vspace{-15pt}
    \label{fig:15Hz_3gflc_1}
\end{figure}

\textbf{\textit{Step 5a:}} \textit{Time-domain analysis:} 
In this step, we compare time-domain simulation results obtained from SPC and QPC models with EMT models. The EMT model of the systems are built in EMTDC/PSCAD \cite{pscad}, which considers averaged models of the converters neglecting dc-side dynamics, distributed parameter representation of the transmission lines (Bergeron model \cite{bergeron_method}), dynamic constant impedance model of the loads, and detailed SG model. A smoothing time constant of $20$ ms is considered for voltage measurement used as feedback signal followed by a $30$ ms first-order delay in the outer voltage control loop shown in Fig.~\ref{fig:gflc_control}(b). It should be noted that, developing such EMT models for multi-IBR-multi-SG systems are significantly more challenging compared to single-IBR-infinite bus models used to validate SSOs in literature \cite{Lingling-InterIBR,Lingling-19-Type4WindModel,Lingling-21-ReducedAnalyticalPV,Lingling-23-NewIBRoscType,SSO_ibr}. This is because the model has to go through various complex startup sequences in the course of reaching the desired equilibrium. This involves further complications when the equilibrium is unstable. Hence in the startup sequence, each GFLC is enabled at distinct time instances, and some of the GFLC parameters such as PI gains of PLLs are initialized with smaller values and ramped up to the targeted values over the first $45$ s to avoid unstable dynamic behaviors of the system even before reaching the equilibrium. Figure~\ref{fig:20Hz_2gflc} presents dynamic responses considering PLL bandwidth of $20$ Hz whereas Fig.~\ref{fig:15Hz_2gflc_2} considers PLL bandwidth of $15$ Hz for the system with $2$-GFLCs (Fig.~\ref{fig:11bus_sys}(a)). Similarly, Figs \ref{fig:20Hz_3gflc} and \ref{fig:15Hz_3gflc_1} showcase the dynamic responses for the $3$-GFLC system (Fig.~\ref{fig:11bus_sys}(b)) considering PLL bandwidths $20$ Hz and $15$ Hz, respectively. 
\vspace{-5pt}

It was noted earlier that for the PLL bandwidth of $20$ Hz, both $2$- and $3$-GFLC configurations are unstable. Hence Figs~\ref{fig:20Hz_2gflc} and \ref{fig:20Hz_3gflc} show how the responses naturally become unstable as simulation time progresses. Due to this fact, the time instances at which the responses from models of different frameworks become unstable are not aligned with each other. Nevertheless, the responses exhibited by each framework appear to be almost similar. Prony analysis \cite{prony} on the $\Delta \omega_{pll,1}$ responses obtained from EMT simulations confirms that the estimated modes closely match those of linearized SPC and QPC models, see Tables~\ref{tab:freq_unstable_mode_2gflc} and \ref{tab:freq_unstable_mode_3gflc}. 

Figures~\ref{fig:15Hz_2gflc_2} and \ref{fig:15Hz_3gflc_1}, present dynamic responses of both system configurations for PLL bandwidth of $15$ Hz following a $1$ s pulse change in $P_{ref}$ of GFLC$1$. 
Once again, dynamic responses produced by models in each framework match  closely except in EMT platform, where the responses are slightly better damped, which is evident from the damping ratios listed in Tables~\ref{tab:freq_unstable_mode_2gflc} and \ref{tab:freq_unstable_mode_3gflc}. Besides this minor difference, the SSO modes were properly captured by all models.

\textbf{\textit{Step 5b:}} \textit{Sensitivity analysis:}
Participation factor analysis suggests that different controller parameters can influence the stability of the SSO modes. 
Based on that we consider certain parameters for sensitivity analysis in frequency domain.
\subsubsection{PLL bandwidth}
\begin{figure}[h!]
    \centerline{
    \includegraphics[width=0.5\textwidth]{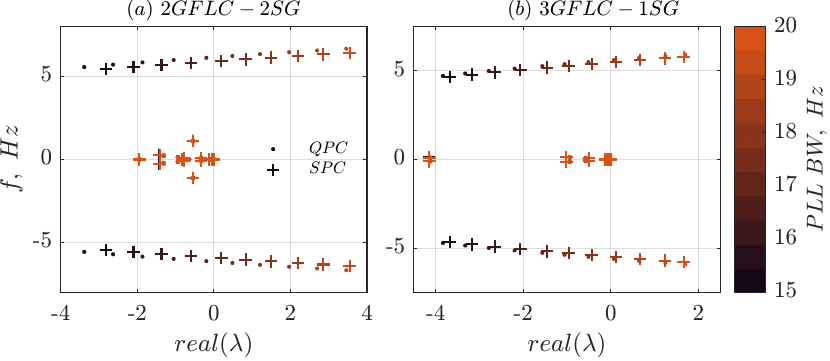}}
    \vspace{-8pt}
    \caption{The locus of the eigenvalues of (a). case ($1$) and (b). case ($2$) as the PLL bandwidth varies.}
    \vspace{-4pt}
    \label{fig:rlocus_bw}
\end{figure}

Figure~\ref{fig:rlocus_bw} shows the root loci of the SSO mode as PLL bandwidth of all GFLCs are simultaneously varied from $20$ Hz to $15$ Hz. It is observed that just below $17$ Hz bandwidth the $2$-GFLC system becomes stable whereas for $3$-GFLC system that bandwidth is about $18$ Hz. It is important to note that, unless otherwise mentioned, only the parameter under investigation is varied and the rest of the parameters remain at original values in these studies. Moreover, different bandwidths of the closed-loop transfer function of the PLL are chosen such that a phase margin of $65.4^{\circ}$ is preserved.
\vspace{-5pt}

\subsubsection{Inner current loop PI gains}
\begin{figure}[h!]
    \centerline{
    \includegraphics[width=0.5\textwidth]{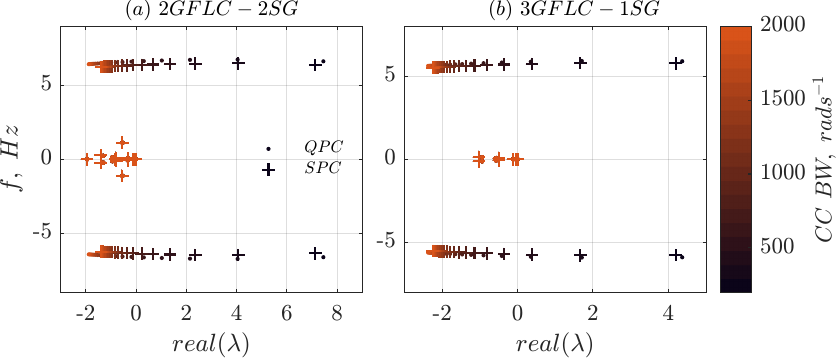}}
    \vspace{-8pt}
    \caption{The locus of the eigenvalues of (a). case ($1$) and (b). case ($2$) as the inner current control PI gains vary. CC: current control}
    \vspace{-6pt}
    \label{fig:rlocus_icl}
\end{figure}

Next, Fig.~\ref{fig:rlocus_icl} illustrates the root loci of the SSO mode as we vary the inner current control loop bandwidth from $200$ to $2,000$ rad/s while maintaining the PLL bandwidth at $20$ Hz.
It is clear that even at $20$ Hz PLL bandwidth, $2$-GFLC and $3$-GFLC systems become stable above $800$ rad/s and $500$ rad/s current controller bandwidths, respectively. 

\subsubsection{Voltage tracking PI gains}
Finally, we investigate how voltage tracking PI gains in reactive power control (i.e., $k_{p,v}$ and $k_{i,v}$ in Fig.~\ref{fig:gflc_control}(b)) impact the system stability. To that end, gains are varied using a multiplier $n$, where $n$ = $1$ corresponds to the nominal values. The unstable SSO mode with $50$\% and $75$\% IBR penetrations are stabilized, respectively at $n$ = $1.45$ and $1.11$.   
Notably, the frequency of the SSO mode varies significantly as the gains are increased. 
\vspace{-5pt}
\begin{figure}[h!]
    \centerline{
    \includegraphics[width=0.5\textwidth]{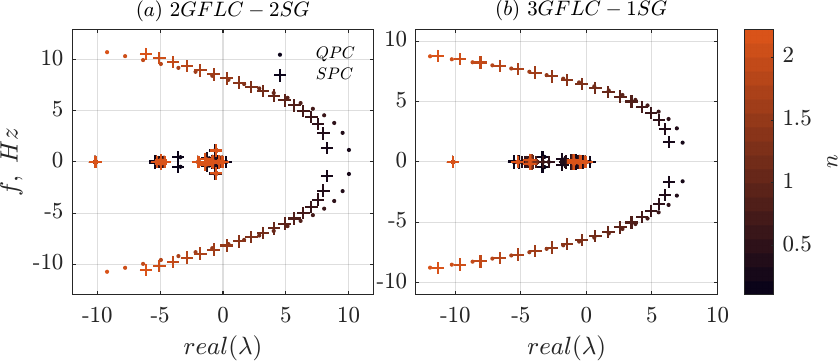}}
    \vspace{-8pt}
    \caption{The locus of the eigenvalues of (a). case ($1$) and (b). case ($2$) as the voltage tracking PI gains in reactive power control vary.}
    \vspace{-15pt}
    \label{fig:rlocus_ovl}
\end{figure}

The above studies indicate that the SSO mode in the SPC model is very closely captured by the QPC model across a wide range of important controller parameters.

\vspace{-5pt}
\section{Conclusion \& Future Work} \label{sec:conclusion}
This paper presented a SPC-based modeling approach in $dq0$ frame for stability analysis of GFLC-related SSOs in multimachine systems under balanced conditions. It was highlighted that the time-varying phasor calculus counterparts of SPC in the forms of baseband-$abc$ and generalized averaging suffer from phasor speed restriction due to the low-pass assumption and approximated representation of models, respectively. On the contrary, the SPC framework does not demonstrate such shortcomings. It was observed that SPC-based models were able to capture SSOs involving GFLCs, which was validated against EMT simulations. 
Moreover, a generalized approach for evaluating modeling adequacy of dynamic systems in small-signal sense was proposed. The proposed method was utilized in assessing adequacy of QPC-based models w.r.t. SPC-based models, where the later consider SG stator, load, and transmission network dynamics while the former represent them algebraically. It was observed that the QPC-based models were also able to capture the SSOs, confirming its potential to be considered for related stability analysis involving GFLCs. Notwithstanding the fact that QPC-based models were deemed adequate in the case studies considered, this may not be true for all configurations of grids involving various types of IBRs. Therefore, the proposed approach for modeling adequacy evaluation should be followed on a case-by-case basis before drawing conclusions. Our ongoing and future research is focused on determining modeling adequacy of systems in presence of GFCs and developing more accurate transmission line models in this context.   

\vspace{-5pt}
\bibliographystyle{IEEEtran}
\bibliography{reference.bib}

\end{document}